\documentstyle[12pt]{article}
\begin{document}
\newcommand{\beqn}{\begin{equation}}
\newcommand{\eeqn}{\end{equation}}
\newcommand{\beqnar}{\begin{eqnarray}}
\newcommand{\eeqnar}{\end{eqnarray}}
\newcounter{abc}
 \newcommand{\beqna}
{\renewcommand{\theequation}{\arabic{section}.\arabic{equation}.\alph{abc}}
     \setcounter{abc}{1}
     \begin{equation}}
\newcommand{\eeqna}{\end{equation}
     \renewcommand{\theequation}{\arabic{section}.\arabic{equation}}}
\newcommand{\beqnb}
{\renewcommand{\theequation}{\arabic{section}.\arabic{equation}.\alph{abc}}
     \setcounter{abc}{2}
     \begin{equation}}
\newcommand{\eeqnb}{\end{equation}
     \renewcommand{\theequation}{\arabic{section}.\arabic{equation}}}
\newcommand{\beqnc}
{\renewcommand{\theequation}{\arabic{section}.\arabic{equation}.\alph{abc}}
     \setcounter{abc}{3}
     \begin{equation}}
\newcommand{\eeqnc}{\end{equation}
     \renewcommand{\theequation}{\arabic{section}.\arabic{equation}}}
\newcommand{\beqnd}
{\renewcommand{\theequation}{\arabic{section}.\arabic{equation}.\alph{abc}}
     \setcounter{abc}{4}
     \begin{equation}}
\newcommand{\eeqnd}{\end{equation}
     \renewcommand{\theequation}{\arabic{section}.\arabic{equation}}}
\newcommand{\beqne}
{\renewcommand{\theequation}{\arabic{section}.\arabic{equation}.\alph{abc}}
     \setcounter{abc}{5}
     \begin{equation}}
\newcommand{\eeqne}{\end{equation}
     \renewcommand{\theequation}{\arabic{section}.\arabic{equation}}}
\newcommand{\seta}{\setcounter{abc}{1}}
\newcommand{\setb}{\setcounter{abc}{2}\addtocounter{equation}{-1}}
\newcommand{\setc}{\setcounter{abc}{3}\addtocounter{equation}{-1}}
\newcommand{\setd}{\setcounter{abc}{4}\addtocounter{equation}{-1}}
\newcommand{\sete}{\setcounter{abc}{5}\addtocounter{equation}{-1}}
\newcommand{\beqnarlett}{
     \renewcommand{\theequation}{\arabic{section}.\arabic{equation}.\alph{abc}}
     \begin{eqnarray}}
\newcommand{\eeqnarlett}{\end{eqnarray}
     \renewcommand{\theequation}{\arabic{section}.\arabic{equation}}}
%
%
\renewcommand{\theequation}{\arabic{section}.\arabic{equation}}
\renewcommand{\thefootnote}{\alph{footnote}}
\setcounter{page}{0}
\pagestyle{empty}
\begin{center}
{\bf \LARGE Spectra of quantum chains \\
	without the Yang-Baxter equation}
\\*[1cm]
{\Large I. Peschel,$^a$   V. Rittenberg$^b$ and U. Schultze$^c$}
\\*[1cm]
\begin{center}
{\it $^a$Max-Planck-Institut f\"ur Physik Komplexer Systeme, \\
Bayreuther Str. 40, Haus 16, 01187 Dresden, Germany \\
Freie Universit\"at Berlin, Fachbereich Physik, \\
Arnimallee 14, 14195 Berlin, Germany (permanent address)
\\*[0.4cm]
$^b$Physikalisches Institut, Universit\"at Bonn, \\
Nussallee 12, 53115 Bonn, Germany
\\*[0.4cm]
$^c$Laboratoire de Physique Th\'{e}orique des Liquides, \\
Universit\'{e} Pierre et Marie Curie, 4 Place Jussieu, \\
75252 Paris, Cedex 05, France}
\end{center}
\vspace{0.6cm}
\begin{abstract}
\vspace{0.6cm}

We study one-dimensional reaction-diffusion models described by master
equations and their associated two-state quantum Hamiltonians. By choosing
appropriate rates, the equations of motion decouple into certain subsets. We
solve the first subset which has a close relation to the problem of lattice
electrons in an electric field. In this way we obtain $L (L - 1) + 1$
energy levels of a quantum chain with $L$ sites. The corresponding
Hamiltonian depends on 7 parameters and does not look integrable using
conventional methods. As an application, we compute the dynamical critical
exponent of a new type of kinetic Ising model.

\end{abstract}
\end{center}

\newpage
\pagestyle{plain}
\section{Introduction}
$\quad {}$ One-dimensional quantum chain Hamiltonian show up in different
contexts in equilibrium
statistical mechanics and many examples of integrable ones are known. They
appear as the logarithmic derivative of the monodromy matrix related to the
Yang-Baxter equation \cite{YB}.
However, there is also a connection to stochastic problems, because a master
equation can be related to a suitable Schr\"odinger equation \cite{KS}. For
example, the diffusion of classical particles with hard-core repulsion on a
lattice leads to the ferromagnetic Heisenberg model \cite{ALEX}, \cite{DI}.
It was realised recently \cite{GWA}, \cite{GSED}, \cite{A},
\cite{E} that in various other stochastic models the corresponding
Hamiltonian is integrable and one can use this integrability in order to
determine the
phase diagram of the system and compute the time evolution of various
average quantities. In the Schr\"odinger equation the wave function is the
time dependent probability distribution, the Hamiltonian is in general
non-hermitian and its matrix elements are constrained by the condition that
the sum of rates for a given process have to add to one (conservation of
probabilities).

In the present paper we adopt the opposite strategy. We ask ourselves if
there are not known examples in non-equilibrium statistical mechanics where
one can solve the master equation or part of it (in a sense to be defined
soon) and thus get in this way the spectrum and wave functions of a
Hamiltonian. We first looked at the Hamiltonian related to the master
equation studied by Kimball \cite{G} and Deker and Haake \cite{Q} where two
wave
functions were known for any number of sites, a few more can be found \cite{DA}
but we were not able to go beyond this point. Another approach which was
proven very powerful is to consider various moments of the probability
destribution (the knowledge of all the moments giving back the probability
distribution) and to find special rates in the master equation such that the
differential equations for various moments decouple into subsets. The first
example of
this kind is the Glauber model \cite{C} used to study the relaxation of the
one-dimensional classical Ising model.

Going to the dual variables description \cite{SI}, the Glauber model
corresponds
to solve the one-spin-string problem (see Sec. 2) which decouples from the
many-spin-strings problem. Subsequently \cite{D}, \cite{T} the many strings
problem
was also solved and implicitely the whole spectrum of the Hamiltonian
derived. The Hamiltonian is however trivial: it corresponds to the $XY$
model in a $Z$ field which can be brought through a Jordan-Wigner
transformation to a bilinear problem in fermionic creation and annihilation
operators and can thus be trivially diagonalized.

Another case is the one considered in Refs. \cite{S}, \cite{R}, \cite{P} and
\cite{O}. This is an
example where the master equation describes chemical processes and brownian
motions \cite{H}. In this case (see Sec. 2 for the terminology) one considers
first processes where two identical molecules $A$ give a molecule $A$ and a
vacancy (coagulation) with a rate "tuned" to the diffusion rate and the
reverse process (decogulation) with an arbitrary rate ($A +$ vacancy
$\to A + A $). This is a one-parameter process since one rate can be used to
fix the time scale. In this case one considers a one-hole probability
function, two-holes etc. ... as moments of the probability function.
It was shown that the equations for the one-hole probability function
decouple from the others. The Hamiltonian in this case is again that of the
$XY$ model in a $Z$ field. Some work was done also for the two-holes
problem \cite{DOE}, \cite{MAB}. Actually, as will be shown in this paper
knowing the
two spin-string solution of \cite{D} and \cite{T} for the Glauber problem
solves the
two-holes problem of the coagulation-decoagulation case. The Glauber
model corresponds to chemical processes in which two molecules annihilate
into two vacancies: $A + A \to $ vacancy $+$ vacancy with a rate tuned to
the diffusion rate and the reverse process with an arbitrary rate. The
investigation of Refs. \cite{S} - \cite{O} has, however, one more case, namely
to the
coagulation-decoagulation processes one adds the birth process (two
vacancies give a molecule $A$ and a vacancy). They were able to solve the
continuum limit of this problem in the case of one-hole. The spectrum can be
expressed in terms of zeroes of one Airy function and the Hamiltonian is not
any more a trivial one. This observation triggered the present paper which is
organized as follows.

In Sec. 2 we give the master equation for a chain of $L$ sites, having on
each site a stochastic variable taking two values. The rates are assumed to
depend on the configuration of two neighbouring sites only. Next we present
the dictionary necessary for chemistry which is also useful in order to
understand our results. The one-dimensional quantum chain associated with
the master equation is also given.

In Sec. 3 we first define the holes and spin-strings. Next we derive the
conditions under which the differential equations for one-hole decouple
from the others. These conditions assure also that the two-holes problem is
decoupled from the three or more holes, etc. since there are five conditions
for the twelve rates. This leaves the Hamiltonian dependent on six parameters
(the normalisation fixes another one). The Hamiltonian one obtains is
non-hermitian. We write the differential equations for the closed ring case
and notice that they depend on one parameter less. This is not the case for
the open chain. We also show that the same differential equations describe
the one-string problem and give the relation between the parameters which
appear in the differential equations and the rates of the one-string problem
(the conditions on the rates are different in this case but their number is
of course the same).

In Sec. 4 we solve the one-hole subset of differential equations for a
finite chain on a closed ring. Their structure depends on one parameter
(called $\delta$ in the text) which is non-negative in the stochastic
problems. If $\delta$ is non-zero, the eigenvalues are given by zeroes of
Lommel polynomials. If $\delta$ is zero, the eigenvalues are of trigonometric
type. We also check under which conditions the spectrum corresponds to a
free fermionic one.

The continuum limit is considered in Sec. 5. If $\delta \neq 0$ one can take
the limit $L \to \infty$ and find that like in the special case of Refs.
\cite{S} - \cite{O}
the eigenvalues are related to zeroes of one Airy function. For $\delta
= 0$, they have a simple expression. The eigenvalues are in general
complex.

As a simple application of our approach, we consider in Sec. 6 the critical
dynamics of the Ising model \cite{SI}, \cite{FOR}, \cite{LAG}. This process is
described by
the same master equation as the one used for diffusion-reaction processes
with supplementary conditions on the rates which assure that for large times
the system reaches the equilibrium distribution of the one-dimensional Ising
model (detailed balance conditions). We have checked whether our seven
parameter Hamiltonian is compatible with the detailed balance conditions
and found two solutions: one is the old Glauber one \cite{C}, \cite{SI},
another one
is new. Its physical signifiance is discussed.

One last comment before going to the bulk of this paper: one-dimensional
diffussion-reaction processes are apparently seen experimentally \cite{KFS}.
This
observation might make the reader have a closer look at the subject.

\section{The Master Equation and 1 - d Quantum Chains}
$\quad {}$
We consider a one-dimensional chain with $L$ sites and on each site $k (k =
1, 2, ... L)$ we take a variable $\beta_k$ which takes two values: $0$ and
$1$. For $\beta_k = 0$ the site $k$ is empty (vacancy), for $\beta_k = 1$
the site $k$ is occupied by a molecule $A$. At $t = 0$ the probability to
find a certain configuration of molecules and vacancies on the chain is
given by the probability distribution
\beqn
P_0 \left( \beta_1, \beta_2, \ldots \beta_L \right) = P_0
\left( \{ \beta \}\right)
\eeqn
If we assume that we have two-body processes only, the time evolution of the
system is given by a master equation:
\begin{eqnarray}
\frac{\partial P \left( \beta_j t \right)}{\partial t} & = &
\sum_{k = 1}^{L - 1} \left[ - V \left( \beta_k, \beta_{k + 1} \right)
\right.
P \left( \beta_1, \ldots , \beta_L ; t \right)  \nonumber \\
& + & \mathop{\sum\nolimits'}\limits^{1}_{{\gamma_k},
{\gamma_{k + 1}} = 0}
\Gamma^{{\gamma_k} {\gamma_{k + 1}}}_{\beta_k, \beta_{k + 1}}
P \left( \beta_1, \ldots , \beta_{k - 1}, \gamma_k, \gamma_{k + 1},
\beta_{k+ 2}, \ldots , \left. \beta_L ; t \right) \right] \nonumber \\
& &
\end{eqnarray}
which gives the probability distribution at the time $t$ for a given
$P (\{\beta \}, t = 0) = P_0$.
The first term on the r. h. s. of (2.2) describes the losses and the second
one the gains.
A prime is used to indicate that in the sum the couple
$\gamma_k = \beta_k , \gamma_{k + 1} = \beta_{k + 1}$ is excluded.
$\Gamma^{\gamma_k , \gamma_{k + 1}}_{ \beta_k , \beta_{k + 1}} $
represents the probability per unit time that the configuration
$(\gamma_k , \gamma_{k + 1})$ on neighbouring sites changes into the
configuration $(\beta_k , \beta_{k + 1})$. The conservation of probabilities
gives:
\begin{eqnarray}
V \left( \gamma_k, \gamma_{k + 1} \right) \; \; = \; \;
\mathop{\sum\nolimits'}\limits^{1}_{\beta_k , \beta_{k + 1} = 0}
\Gamma^{\gamma_k , \gamma_{k + 1}}_{\beta_k , \beta_{k + 1}}
\end{eqnarray}
where again the prime in the sum indicates that the couple
$\gamma_k = \beta_k , \gamma_{k + 1} = \beta_{k + 1}$ is excluded. The
master equation (2.2) is quite general and can have various physical
interpretations (in Sec. 6 we will apply is to study critical dynamics), in
the present section however we use the language of diffusion-reactions
processes where the rates have a simple meaning.
The following processes are included in the master equation (with
$0$ a vacancy $(\beta = 0)$ and $A$ a molecule $(\beta = 1 )$)
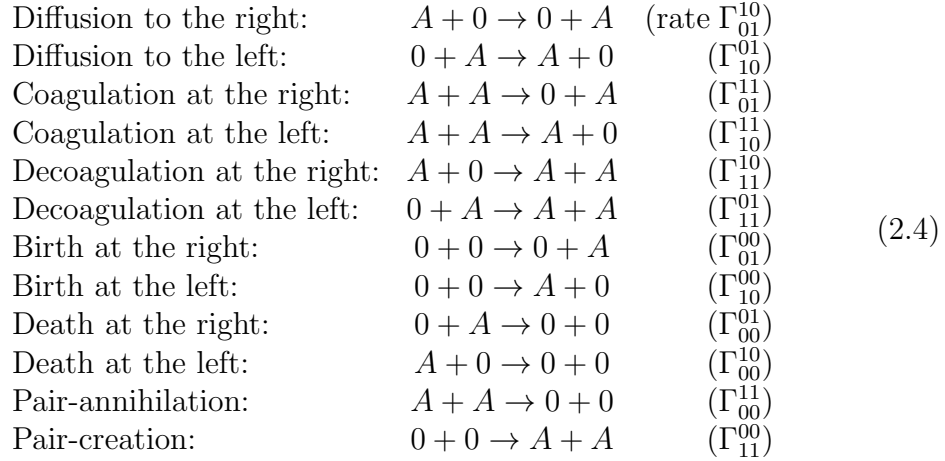
\beqn
\begin{array}{lcr}
{\mbox {Diffusion to the right:}} & A + 0 \to 0 + A & ({\mbox{rate}}\;
\Gamma^{10}_{01}) \nonumber \\
{\mbox {Diffusion to the left:}} & 0 + A \to A + 0 & (\Gamma^{01}_{10})
\nonumber \\
{\mbox {Coagulation at the right:}} & A + A \to 0 + A & (\Gamma^{11}_{01})
\nonumber \\
{\mbox {Coagulation at the left:}} & A + A \to A + 0 & (\Gamma^{11}_{10})
\nonumber \\
{\mbox {Decoagulation at the right:}} & A + 0 \to A + A & (\Gamma^{10}_{11})
\nonumber \\
{\mbox {Decoagulation at the left:}} & 0 + A \to A + A & (\Gamma^{01}_{11})
\nonumber \\
{\mbox {Birth at the right:}} & 0 + 0 \to 0 + A & (\Gamma^{00}_{01}) \nonumber
\\
{\mbox {Birth at the left:}} & 0 + 0 \to A + 0 & (\Gamma^{00}_{10}) \nonumber
\\
{\mbox {Death at the right:}} & 0 + A \to 0 + 0 & (\Gamma^{01}_{00}) \nonumber
\\
{\mbox {Death at the left:}} & A + 0 \to 0 + 0 & (\Gamma^{10}_{00}) \nonumber
\\
{\mbox {Pair-annihilation:}}	      & A + A \to 0 + 0 & (\Gamma^{11}_{00})
\nonumber \\
{\mbox {Pair-creation:}}	      & 0 + 0 \to A + A & (\Gamma^{00}_{11})
\end{array}
\end{equation}

In Ref. \cite{A} (see also the references included therein), several special
cases have been considered in detail. The notations for the rates used in
Ref. \cite{A} are related to those used in the present paper:
\beqn
\Gamma^{\gamma , \delta}_{\alpha, \beta} = w_{\gamma - \alpha, \delta -\beta}
(\alpha , \beta) \; \; ; \; \; V (\alpha ,\beta) = w_{00} (\alpha ,\beta)
\eeqn
To a given master equation one can associate \cite{KS} a Schr\"odinger equation
\beqn
\frac{\partial}{\partial t} \mid P > \; = \; - H \mid P >
\eeqn
as follows. Take a $2^L$-dimensional orthogonal basis:
\beqn
\mid \{\beta \} > = \mid \beta_1 , \beta_2 \ldots , \beta_L >\; , \;
< \{ \beta^\prime \} \mid \{ \beta \} > =
\delta_{\{ \beta \}, \{\beta^\prime\}}
\eeqn
and denote:
\beqn
\mid P > \; = \; \sum_\beta P( \{ \beta\} ; t) \mid \{ \beta \} > \; ; \;
\mid P_0 > \; = \; \sum_\beta P_0( \{ \beta\}) \mid \{ \beta \} >
\eeqn
In order to write the Hamiltonian, we first define a basis in the space of
$2 \times 2$ matrices taking
$E^{\alpha , \beta} (\alpha, \beta = 0, 1)$
with matrix elements:
\beqn
(E^{\alpha , \beta})_{\gamma , \delta} = \delta_{\alpha, \gamma} \; \;
\delta_{\beta , \delta}
\eeqn
On each site $k$ of the chain we take the matrices
$E^{\alpha , \beta}_k (k = 1, 2, \ldots L)$. It was shown in Ref. \cite{A} that
in this basis the Hamiltonian $H$ occuring in Eq. (2.6) has the following
expression:
\beqn
H = \sum^{L - 1}_{k = 1} H_k
\eeqn
where
\beqn
H_k = - \sum T^{\alpha , \beta}_{\gamma , \delta} \; E^{\gamma \alpha}_k
\; E^{\delta \beta}_{k + 1}
\eeqn
with
\begin{eqnarray}
T^{\alpha , \beta}_{\gamma , \delta} & = &
\Gamma^{\alpha , \beta}_{\gamma , \delta} \; (\alpha = \gamma , \; \beta =
\delta \; \; {\mbox{excluded}})
 \\
T^{\alpha , \beta}_{\alpha , \beta} & = & - V (\alpha , \beta)
\end{eqnarray}
where we have used Eq. (2.3). The Hamiltonian $H$ is in general
non-hermitian and due to the probability conservation (Eq. (2.3)), it
satisfies the remarkable relation:
\beqn
< 0 \mid H = 0
\eeqn
where the bra ground-state $< 0 \mid$ is
\beqn
< 0 \mid = \sum_{\{ \beta \}} < \{ \beta \} \mid \; = \;
\left\langle {1 \choose 1} \otimes  \right.
{1 \choose 1} \otimes \ldots \otimes {1 \choose 1}  \mid
\eeqn
{}From Eq. (2.14) it follows that the ground-state energy is zero. The
Hamiltonian might have more than one eigenstate of zero energy i. e. there
may be more than one stationary solution.

It is instructive to write the $4 \times 4$ matrix given by Eq. (2.11)
explicitely:
\beqn
H_k = \left(
\begin{array}{cccc}
\Gamma^{00}_{01}  +  \Gamma^{00}_{10} + \Gamma^{00}_{11}
& -  \Gamma^{01}_{00} &	 -  \Gamma^{10}_{00} 	& -  \Gamma^{11}_{00}
\nonumber \\[0.5ex]
 -  \Gamma^{00}_{01} & \Gamma^{01}_{00}  +  \Gamma^{01}_{10} +
\Gamma^{01}_{11} & -  \Gamma^{10}_{01} & - \Gamma^{11}_{01}
\nonumber \\[0.5ex]
 -  \Gamma^{00}_{10} 	& -  \Gamma^{01}_{10} &
\Gamma^{10}_{00}  +  \Gamma^{10}_{01} + \Gamma^{10}_{11}
& - \Gamma^{11}_{10}
\nonumber \\[0.5ex]
 -  \Gamma^{00}_{11} & -  \Gamma^{01}_{11} & -  \Gamma^{10}_{11} &
\Gamma^{11}_{00}  +  \Gamma^{11}_{01} + \Gamma^{11}_{10} \nonumber \\[0.5ex]
& & &
\end{array}
 \right)
\eeqn
Notice that the sum of each column in (2.16) vanishes, this is a consequence
of Eq. (2.14).

Let us now consider an observable
$X \left( \beta_1 , \ldots \beta_L \right) = X (\{ \beta \})$, its average
quantity can be computed as follows:
\begin{eqnarray}
< X > ( t ) & = & \sum_{\{ \beta \}} X (\{ \beta \} ) P ( \{ \beta \}, t)
\nonumber \\
& = & < 0 \mid X \mid P > = < 0 \mid X e^{-H t} \mid P_0 >
\end{eqnarray}
The time derivative of $< X > (t)$ is obviously:
\begin{equation}
\frac{d < X > (t)}{dt} \; = \; - < 0 \mid X H e^{-H t} \mid P_0 > \; = \;
< 0 \mid [H , X] e^{-H t} \mid P_0 >
\end{equation}
where we have used Eq. (2.14). In the next Section two types of observables
will be used. The first one is a hole of length $n$ starting at the site
$j$:
\begin{equation}
X = \prod^{j + n - 1}_{k = j} \delta (\beta_k)
\end{equation}
Its average is:
\begin{eqnarray}
K ( j, n + j - 1 ) & = & \sum_\beta \; \delta (\beta_j ) \delta (\beta_{j + 1})
\ldots \delta (\beta_{j + n - 1}) P (\{\beta\} ; t)
\nonumber \\
& = & < 0 \mid \; \prod^{n + j - 1}_{k = j} \; E^{00}_{k} e^{-H t}  \; \mid
P_0 >
\end{eqnarray}
where we have used Eq. (2.9). The time dependent function
$ K (j, n + j - 1)$
gives the probability that $n$ consecutive sites starting with the site
$j$, are empty (a hole). This observable is sometimes used also in
equilibrium problems \cite{KIEU}. A second observable which will be of interest
is a spin-string of length $n$ starting at the site $j$:
\begin{equation}
X = \prod^{j + n - 1}_{k = j}
\left( \delta (\beta_k) - \delta (\beta_k + 1) \right)
\end{equation}
Its average quantity is:
\begin{equation}
S ( j, n + j - 1) = < 0 \mid \prod^{n + j - 1}_{k = j} \; \sigma^z_k \; \;
e^{-H t} \mid P_0 >
\end{equation}
where
\begin{equation}
\sigma^z = {1 \; \; 0 \choose 0  - 1} = E^{00} - E^{11}
\end{equation}
Actually one can define an observable interpolating between the one-hole and
the one-spin-string taking
\beqn
X = \prod^{j + n - 1}_{k = j}
\left( \frac{1}{1 + \xi}  \delta (\beta_k) - \right.
\left. \frac{\xi}{1 + \xi} \delta (\beta_k + 1) \right)
\eeqn
where $\xi$ is a parameter. Its average quantitiy being:
\beqn
X (j , n + j - 1) =
< 0 \mid \prod^{n + j - 1}_{k = j} \; \tau_k \; \; e ^{-H t} \; \mid P_0 >
\eeqn
where
\beqn
\tau = \frac{1}{1 + \xi} E^{00} - \frac{\xi}{1 + \xi} E^{11}
\eeqn
\section{Decoupling of the master equation into subsets}
\setcounter{equation}{0}
$\quad {}$
As we have seen in the proceeding Section, the master equation (2.2) is
equi\-valent to the Schr\"odinger equation (2.6). Thus finding the solution of
the $2^L$ linear differential equations (2.2) will give the eigenvalues and
eigen-functions of the Hamiltonian given by the Eqs. (2.10) - (2.11). The
main idea that we are going to use in this paper is to find conditions under
which the $2^L$ differential equations decouple into subsets which can be
solved independently. This corresponds to bringing the Hamiltonian to a
block-diagonal form. There are some obvious sub-sets that one might
consider:
\begin{itemize}
\item[a)]
\underline{Holes}
The definition of a hole was given by Eq. (2.20). One can
ask for conditions on the rates such that the one-hole differential
equations decouple from the others. The two-holes equations decouple from
more than two-holes etc.... Since obviously knowing the solution of all the
holes probabilities gives the probability distribution $P (\{ \beta \} ,
t)$, this corresponds to the diagonalisation of the Hamiltonian.
\item[b)]
\underline{Spin-strings}
The definition of a spin-string was given by Eq.
(2.22). One can consider the hierarchy of the sub-sets of one string, two
strings etc. Again knowing all the spin-strings functions is equivalent to
the knowledge of $P (\{ \beta \}, t) $ (the proof is trivial).
\item[c)]
\underline{Vacancies}
The $N$-vacancies function is
\begin{eqnarray}
V ( i_1, i_2, \ldots i_N ) & = & \sum_\beta \delta (\beta_{i_1}) \; \delta
(\beta_{i{_2}}) \ldots \delta (\beta_{i_N}) P(\{ \beta \} ; t) \nonumber \\
& = & < 0 \mid E^{00}_{i_1} E^{00}_{i_2} \ldots E^{00}_{i_N} \; \; e^{-H t}
\mid P_0 >
\end{eqnarray}
The hierarchy is one vacancy, two vacancies etc. ...
(see also \cite{M}, \cite{N} \cite{GS}).
\item[d)]
\underline{Spins}
The $N$-spins function is
\begin{eqnarray}
S ( i_1 , i_2, \ldots , i_N ) & = & \sum_\beta (\delta (\beta_{i{_1}}) -
\delta (\beta_{i{_2}})) \ldots ( \delta (\beta_{i{_N}}) -
\delta (\beta_{i{_N + 1}})) P ( \{ \beta \} ; t)
\nonumber \\
& = & < 0 \mid \sigma^z_{i{_1}} \; \sigma^z_{i{_2}} \ldots \sigma^z_{i{_N}}
{}^{- H t}
\mid P_0 >
\end{eqnarray}
with the hierarchy one spin, two spins etc. ...
\end{itemize}
As we are going to show soon, once we solve the holes case we can get
through a similarity transformation the spin-strings solution.

It is interesting at this point to solve a small counting problem. Take first
an open chain with $L$ sites and count how many kind of holes $N_m$
(various lengths and positions)
one can have for a given total of $m$ holes
$(m = 1, 2, . , \frac{L}{2}$ for $L$ even,
$m = 1, 2, \ldots \frac{L + 1}{2}$ for $L$ odd). The answer is:
\begin{equation}
N_m = \frac{( L + 1 ) !}{( 2 m ) ! ( L - 2 m + 1) !} = C^{2 m}_L  \;
+ \; C^{2 m - 1}_L
\end{equation}

The meaning of the right hand side of Eq. (3.3) will be apparent
immediately. Consider now for example the following Hamiltonian:
$$
\begin{array}{lcl}
H  & = & \sum\limits^{L - 1}_{i = 1} \; H_i  \hspace{9.7cm} ({\rm 3.4.a})
\\[3ex]
{}
\\
H_i & = & - \frac{\eta}{2}
\left[
\eta \sigma^x_i \sigma^x_{i + 1} \; + \;
\eta^{-1} \sigma^y_i \sigma^y_i \; + \;
\sigma^z_i + \sigma^z_{i + 1} - \eta -1
\right] \hspace{2.9cm}  ({\rm 3.4.b})
\end{array}
$$
where $\eta$ is a free parameter and $\sigma^x , \sigma^y$ and $\sigma^z$
are Pauli matrices. This Hamiltonian will correspond to a special choice of
the parameters in our solution of the decoupling problem. Here we are only
interested in the structure of the problem. The Hamiltonian (3.4) can be
diagonalized through a Jordan-Wigner transformation \cite{K}. The result for
the
open chain has the form:
\begin{equation}
\setcounter{equation}{5}
H = \sum^L_{k = 1} \varepsilon_k n_k
\end{equation}
where $n_k$ are fermionic occupation numbers $(n_k = 0, 1)$ and
$\varepsilon_1 = 0$ (there is a zero fermionic mode). If one computes now
the number of states in the sectors with $2m$ and $(2m - 1)$ fermions
one finds precisely (3.3). The fermionic vacuum does not appear in the holes
calculation. The reason is simple: in a probabilistic calculation one does
not need to know $2^L$ probabilities $P ( \beta_1, \ldots , \beta_L)$ but only
$2^L - 1$ since the sum of probabilities adds to one.

We now consider a closed ring with $L$ sites. In this case the number of
holes in the subset with $m$-holes $(m = 1, 2, \ldots , \frac{L}{2}$ for
$L$ even, $m = 1, 2, \ldots , \frac{L - 1}{2}$ for $L$ odd) is
\begin{equation}
N_m = \frac{2 L !}{(2m) ! (L - 2m)!}
\end{equation}
except for $m = 1$ where we have one more hole than one obtains from Eq.
(3.6). This special hole corresponds to the situation where all the $L$
sites are empty. The reason for this choice of the supplementary hole comes
from a decoupling of this hole from the other functions in the 1-hole subset
(see Sec. 4). Adding all states coming from all the $m$-holes subsets one gets
$2^L - 1$ as one should (one more state comes from probabilities
conservation).

We now return to the Hamiltonian (3.4) with periodic boundary conditions
which can be diagonalized through a Jordan-Wigner transformation \cite{BCD},
\cite{I} and its spectrum can be obtained taking the even number of fermions
sector of two free-fermionic Hamiltonians, one with periodic boundary
conditions and the other with antiperiodic boundary conditions. Notice that
both Hamiltonians have a zero ground state energy. The number of states in
the $2m$ fermions sector of both Hamiltonians give together $N_m$ of Eq.
(3.6). We are left with two fermionic vacua. One corresponds to the total
empty one-hole the other to the redundand state (the probabilities add to
one). This calculation suggests that for special choices of the parameters of
our problem we will get Hamiltonians which correspond to free fermions. This
will be proven to be indeed the case. We will also find an unexpected
phenomenon, where the closed chain problem will be described by free fermions
but not the open chain. We will also get that in general the spectrum is not
given by free fermions at all, but we believe that our counting could help
clarify the algebraic structure of the problem.

The vacancies problem will not be considered here but can be found in
\cite{GS}. One finds a non-hermitian Hamiltonian depending on ten parameters
whose spectrum is given by the $XXZ$ Heisenberg chain in a $Z$ field.
The same similarity transformation used for the holes-spin-strings
correspondence can be used to solve the spins problem once the vacancies
solution is known.

We now start with the holes problem and ask for the
conditions on the rates such that the time derivative of a one hole
probability function
\begin{equation}
\frac{d K (j, j + n - 1 )}{dt} = < 0 \mid
\left[ H, \prod^{j + n - 1}_{k = j } \; \; E^{00}_k \right] \; e^{-H t} \mid
P_0 >
\end{equation}
can be expressed in terms of one-hole probability functions only. A
straightforward calculation leads to the following five conditions:
\begin{equation}
\Gamma^{11}_{00} \; = \; \Gamma^{01}_{00} \; = \; \Gamma^{10}_{00} \; = \; 0
; \quad
\Gamma^{11}_{10} \; = \; \Gamma^{01}_{10} \; ; \;
\Gamma^{10}_{01} \; = \; \Gamma^{11}_{01}
\end{equation}
The physical meaning of the conditions (3.8) is the following. All processes
creating $A$ molecules (creation, birth and decoagulation) are allowed.
{}From the processes which "kill" molecules only coagulation processes are
allowed and they are tuned to the diffusion rates: the rate of coagulation
at the right (left) is equal to the diffusion rate to the right (left). To
shorten the notations, we will denote:
\begin{eqnarray}
a_R & = & \Gamma^{10}_{01} = \Gamma^{11}_{01} ; \; \;
a_L = \Gamma^{01}_{10} = \Gamma^{11}_{10}
\nonumber \\
d_R & = & \Gamma^{10}_{11} , d_L = \Gamma^{01}_{11} \; \; , \; \;
b_R = \Gamma^{00}_{01} \; , \; b_L = \Gamma^{00}_{10} \; , \;  c =
\Gamma^{00}_{11}
\end{eqnarray}
Thus the $a$'s describe diffusion and coagulation, the $d$'s decoagulation,
the $b$'s birth and $c$ pair-creation.
If we take into account Eqs. (3.9), the $4 \times 4$ matrix $H_k$ of Eq.
(2.16) becomes:
\beqn
H_k = \left(
\begin{array}{cccc}
c + b_L + b_R & 0 & 0 & 0 \nonumber \\
- b_R & a_L + d_L & - a_R & - a_R \nonumber \\
- b_L & - a_L & a_R + d_R & - a_L \nonumber \\
- c_L & - d_L & - d_R & a_L + a_R
\end{array}
\right)
\eeqn
and it is the Hamiltonian (2.10) with (3.10) that we want to diagonalize.
This Hamiltonian is non-hermitian and depends on seven parameters, one of
which will fix the time scale. It is now useful to use the chemical picture
given by Eq. (2.4) in order to look for global symmetries of the
Hamiltonian. In general one finds none. All one can say is that if the rates
are all positive, all the eigenvalues are non-negative.

The differential equations derived from Eq. (3.7) for the one-hole
probability functions for a chain with $L$ sites depend on the boundary
conditions:
\underline{Closed ring, $L (L - 1) + 1$ equations}
\begin{eqnarray}
\frac{d K (j , j + n - 1)}{dt} & = & -  [\gamma + (n - 1) \delta ]
K (j , j + n - 1) \;  \nonumber \\
& {} & + a_L  \;  K (j + 1 , j + n - 1)
\nonumber \\
& + & a_R K (j , j + n - 2) \; + \; \alpha_L K (j , j + n)
\nonumber \\
&  + & \alpha_R K (j - 1 , j + n - 1) \; \; (1 \leq n \; \leq L - 1) \\
\frac{d K (j , j + L - 1)}{dt} & = & - L \delta K (j , j + L - 1) \quad
\quad (n = L)
\end{eqnarray}
with the notation:
\begin{equation}
K (j , j) = 1 \quad \quad (n = 0)
\end{equation}
Obviously $K (j , j + L - 1)$ is $j$ independent.
\\
\underline{The open chain, $L (L + 1) /2$ equations.}

Equation (3.11) stays unchanged except when $j = 1$ or $j = L - n + 1$
when we have:
\begin{eqnarray}
\frac{d K ( 1 , n)}{dt} & = &
- [\gamma_L + (n - 1)\delta ] K (1 , n ) + a_R K ( 1, n - 1) \nonumber \\
&& + \alpha_L K ( 1, n + 1 )
\end{eqnarray}
\begin{eqnarray}
\frac{d K ( j , L )}{dt} & = &
- [\gamma_R + (n - 1) \delta ] K (j , L) + a_L K (j + 1 , L) \nonumber \\
&& + \alpha_R K (j - 1 , L)
\end{eqnarray}
Equation (3.12) is replaced by:
\begin{equation}
\frac{d K ( 1, L)}{dt} = - ( L - 1 ) \delta K ( 1 , L )
\end{equation}
and the notation (3.13) used in Eqs. (3.11), (3.14) and (3.15) stays
unchanged. In the above Equations we have denoted:
\begin{eqnarray}
\alpha_R & = & a_R + d_R - b_R - c \; = \; \Gamma^{10}_{01} + \Gamma^{10}_{11}
- \Gamma^{00}_{11} - \Gamma^{00}_{01} \nonumber \\
\alpha_L & = & a_L + d_L - b_L - c \; = \; \Gamma^{01}_{10} + \Gamma^{01}_{11}
- \Gamma^{00}_{11} - \Gamma^{00}_{10} \nonumber \\
a_R & = & \Gamma^{10}_{01} , \quad  a_L = \Gamma^{01}_{10} \nonumber \\
\gamma_R & = & a_L + a_R + d_R = \Gamma^{01}_{10} + \Gamma^{10}_{01}
+ \Gamma^{10}_{11}  \nonumber \\
\gamma_L & = & a_L + a_R + d_L = \Gamma^{01}_{10} + \Gamma^{10}_{01}
+ \Gamma^{01}_{11} \nonumber \\
\gamma & = & \gamma_R + \gamma_L \nonumber \\
\delta & = & b_R + b_L + c = \Gamma^{00}_{01} + \Gamma^{00}_{10}
+ \Gamma^{00}_{11}
\end{eqnarray}
Notice that the equations for the open chain depend on seven parameters
($\alpha_R , \alpha_L a_R , a_L , \gamma_R , \gamma_L $ and $\delta$). In
the case of the ring however, the differential equations depend only on six
parameters since $\gamma_R$ and $\gamma_L$ appear in the combination
$\gamma_R + \gamma_L$ only. One also sees that the time derivative of the
pro\-bability to find a hole of length $n$ is related to the probability
functions for one-hole of length $n - 1 , n$ and $n + 1$ .

We now consider the two-holes probability function:

\begin{equation}
K ( j, n + j - 1 \; ; \; k , k + m - 1 ) = < 0 \mid \prod^{n + j - 1}_{r = j}
\; E^{00}_k \prod^{j + m - 1}_{s = k} \; E^{00}_s \;e^{- H t} \; \mid P_0 >
\end{equation}
which describes the probability of having a hole of length $n$
starting on the site $j$ and a second hole of length $m$ starting on the
site $k$. A straighforward calculation gives the following differential
equations for the two-holes probability function in the case of the ring
geometry:
\begin{eqnarray*}
\frac{d K (j , j + n - 1 ; k , k + m - 1 )}{d t} & = & \qquad {}
\end{eqnarray*}
\begin{eqnarray}
& - &  [2 \gamma + (m + n - 2) \delta] \: K ( j , j + n - 1 ; k , k + m - 1 )
\nonumber \\
& + & a_L [ K ( j + 1 , j + n - 1 ; k , k + m - 1 ) +
K (j , j + n - 1 ; k + 1 , k + m - 1 ) ]
\nonumber \\
 & + & a_R [ K (j , j + n - 2 ; k , k + m - 1 ) + K (j , j + n - 1 ; k , k + m
- 2)
] \nonumber \\
& + & \alpha_L [ K (j , j + n ; k , k + m - 1 ) + K (j , j + n - 1 ; k , k + m
) ]
\nonumber \\
& +  &\alpha_R [ K ( j - 1 , j + n - 1 ; k , k + m - 1 ) + K (j , j + n - 1 ; k
- 1 , k + m - 1)]
\nonumber\\
& & \nonumber\\
\end{eqnarray}
As one notices the conditions (3.8) which decouple the one-hole problem from
more holes, decouples the two-holes problem from three and more holes. The
two-holes problem contains one-hole probability functions. This happens when
functions of the type
\begin{equation}
K (j , j + n - 1 ; j + n , k + m - 1 ) = K (j , j + m + n - 1 )
\end{equation}
occur in the right hand side of Eqs. (3.19). The same pattern occurs in the
problem of three or more holes and thus the conditions (3.8) have produced a
decoupling into sub-sets of the master-equation. The same happens for the
open chain.

We now turn to the decoupling problem in the case of spin-strings. The
one-spin string is defined by Eq. (2.22), the probability for
two-spin-strings is:
\begin{equation}
S (j , j + n - 1 ; k , k + m - 1 ) = < 0 \mid \prod^{j + n - 1}_{r = j} \;
\sigma^z_r \; \prod^{k + m - 1}_{s = k} \; \sigma^z_s \; e^{- H t} \mid P_0 >
\end{equation}
etc. ... Let us show that the decoupling problem for the spin-strings takes
us to the same differential equations as the ones for the holes with a
simple replacement $K \to S$. The expressions of the constants
$a_{R , L} , b_{R , L } , d_{R , L} $ and $c$ respectively
$\alpha_{R , L} , \gamma_{R , L}$ and $\delta$ in terms of rates have to be
changed.

Consider two matrices $b$ and $b^{- 1}$:
\begin{equation}
b = { 1  -1 \choose 0 \; \; 2} \; \; , \; \; b^{-1} =
{1 \; \; 1/2 \choose 0 \; \; 1/2}
\end{equation}
and correspondingly the matrices:
\begin{equation}
B = \prod^L_{k = 1} b_k \; \; , \; \; B^{-1} = \prod^L_{k = 1} b^{- 1}_k
\end{equation}
we perform a similarity transformation \cite{I} which takes us from the
Hamiltonian $H$ given by Eq. (2.10) with $H_k$ given by Eq. (3.10) to $H^S$
respectively $H^S_k$:
\begin{equation}
H^S = B^{- 1} H B
\end{equation}
and
\newpage
\[
H^S_k = \left(
\begin{array}{r}
\frac{b_R + b_L}{2} + \frac{3 c}{4 } \; \;
- \frac{(b_R + b_L)}{2} - \frac{3 c}{4 } + \frac{d_L}{2}
\nonumber \\
- \left( \frac{b_R}{2} + \frac{c}{4} \right)  \quad \quad
\frac{b_R + d_L}{2} + a_L + \frac{c}{4}
\nonumber \\
- \left( \frac{b_L}{2} + \frac{c}{4} \right) \quad  \quad
\frac{b_L - d_L}{2} - a_L + \frac{c}{4}
\nonumber \\
- \frac{c}{4}  \qquad \qquad \quad
- \frac{d_L}{2} + \frac{c}{4}
\end{array}
\qquad \qquad \qquad \qquad \qquad \qquad \qquad \qquad
\right.
\]
\[
\qquad \qquad \qquad
\left.
\begin{array}{l}
- \frac{(b_R + b_L)}{2} + \frac{d_R}{2} - \frac{3 c}{4 } \; \; \;
\frac{(b_R + b_L)}{2} - \frac{(d_R + d_L)}{2} - a_L - a_R + \frac{3 c}{4}
\nonumber \\
\frac{b_R - d_R}{2} - a_R + \frac{c}{4}  \qquad \;
- \frac{b_R }{2} + \frac{d_R - d_L}{2} - \frac{c}{4}
\nonumber \\
\frac{b_L + d_R}{2} + a_R + \frac{c}{4}  \qquad \;
- \frac{b_L }{2} + \frac{d_L - d_R}{2} - \frac{c}{4}
\nonumber \\
- \frac{d_R}{2} + \frac{c}{4}  \qquad \qquad \quad \;
\frac{d_R + d_L}{2} + a_L + a_R - \frac{c}{4}
\end{array}
\right)
\]
\beqn
{}
\eeqn
Notice that as in Eq. (2.16) the sum of the matrix elements in each column
are zero and consequently the matrix elements of (3.25) can be interpreted
as rates provided that they are positive, we will denote them by
$\tilde{\Gamma}^{\gamma , \delta}_{\alpha , \beta}$.
Instead of the five conditions (3.8) on the rates
$\Gamma^{\gamma , \delta}_{\alpha, \beta}$ we now find five different
conditions on the rates $\tilde{\Gamma}^{\gamma , \delta}_{\alpha, \beta}$:
\begin{eqnarray}
\tilde{\Gamma}^{11}_{00} + \tilde{\Gamma}^{00}_{11} \; = \;
\tilde{\Gamma}^{01}_{10} + \tilde{\Gamma}^{10}_{01} & ; & {} \nonumber \\
\tilde{\Gamma}^{10}_{00} + \tilde{\Gamma}^{10}_{11} \; = \;
\tilde{\Gamma}^{00}_{10} + \tilde{\Gamma}^{00}_{01} & ; &
\tilde{\Gamma}^{01}_{00} + \tilde{\Gamma}^{01}_{11} \; = \;
\tilde{\Gamma}^{00}_{10} + \tilde{\Gamma}^{00}_{01} \nonumber \\
\tilde{\Gamma}^{11}_{01} + \tilde{\Gamma}^{10}_{11} \; = \;
\tilde{\Gamma}^{01}_{11} + \tilde{\Gamma}^{00}_{01} & ; &
\tilde{\Gamma}^{11}_{10} + \tilde{\Gamma}^{01}_{11} \; = \;
\tilde{\Gamma}^{00}_{10} + \tilde{\Gamma}^{10}_{11}
\end{eqnarray}
The spectrum of the Hamiltonian (3.25) is again non-negative if the
$\tilde{\Gamma}^{\gamma  \delta}_{\alpha  \beta}$ are positive. This
implies a different domain for the parameters
$a_{R , L} , b_{R , L} , d_{R , L}$
and $c$ than the one obtained from the positivity of the rates
$\Gamma^{\gamma \delta}_{\alpha \beta}$
occuring in Eq. (3.10).

Let us now consider the one-hole function (2.19) and make use of the
similarity transformation (3.24):
\begin{equation}
K (j , n + j - 1) \; = \; < 0 \mid \; \prod^{j + n - 1}_{k = j} \; E^{00}_k B
e^{-
{H^S} t}  B^{-1} \; \mid P_0 >
\end{equation}
Let us now notice (see Eq. (2.15)) that
\begin{equation}
< 0 \mid b \; = \; < 0 \mid
\end{equation}
\begin{equation}
< 0 \mid E^{00}_k B \; = \; < 0 \mid (E^{00}_k - E^{01}_k) \; = \; < 0 \mid
(E^{00}_k - E^{11}_k) = < 0 \mid \sigma^z_k
\end{equation}
With (3.28) and (3.29) we get instead of Eq. (3.27)
\begin{equation}
K (j , n + j - 1) = < 0 \mid \; \prod^{j + n - j}_{k = j} \; \sigma^Z_k
e^{-{H^S} t} \; \mid P_0^S >
\end{equation}
where
\begin{equation}
\mid P_0^S > \: = \; B^{-1} \mid P_0 >
\end{equation}
Comparing the Eqs. (2.22) and (3.30) we learn that the one-hole probability
function at time $t$ given by a Hamiltonian $H$ with an initial condition
$P _0 (\{ \beta \})$ coincides with the one-spin-string function given by
the Hamiltonian $H^S$ and initial condition $P^S_0 (\{ \beta \})$. The rates
$\Gamma^{\gamma \delta}_{\alpha \beta}$ and
$\tilde{\Gamma}^{\gamma \delta}_{\alpha \beta}$ are both dependent on the
seven parameters $a_{R, L} , b_{R, L} , d_{R , L}$ and $c$ but as we
mentioned the domain of the parameters which makes the rates
$\Gamma^{\gamma \delta}_{\alpha \beta}$ positive is in general different
from
the domain where $\tilde{\Gamma}^{\gamma \delta}_{\alpha \beta}$ are positive.
The
differential equations satisfied by the spin-string functions which depend on
the parameters $a_{R, L} , \alpha_{R, L} , \gamma_{R , L}$ and $\delta$ are
identical with those satisfied by the holes probability functions, the
expression of these parameters in terms of the rates
$\tilde{\Gamma}^{\gamma \delta}_{\alpha \beta}$ however, differs from
(3.17) where the ${\Gamma}^{\gamma \delta}_{\alpha \beta}$ rates were used.
For example:
\begin{equation}
\delta = 2 (\tilde{\Gamma}^{00}_{10} + \tilde{\Gamma}^{00}_{01})  \; ; \;
\gamma = 2 (\tilde{\Gamma}^{01}_{10} + \tilde{\Gamma}^{10}_{01} +
\tilde{\Gamma}^{00}_{10} + \tilde{\Gamma}^{00}_{01})
\end{equation}
Notice that $\gamma$ and $\delta$ are non-negative as for the rates
$\Gamma^{\gamma \delta}_{\alpha \beta}$ (see Eq. (3.17)).
Obviously the
transformation (3.23) takes the two-holes probability functions into the
two-spin-string functions etc.... The same similarity transformation takes
vacancies (Eq. (3.1)) into spins (Eq. (3.2)).
Finally let us note that the more general transformation
\begin{equation}
b =
\left( {\frac{1}{1 + \xi} \atop 0}  {\frac{- \xi}{1 + \xi} \atop 1} \right)
\end{equation}
takes the one-hole function into the more general observable (2.25).
This observation
is relevant for two reasons. On one side it shows that more general
observables with the same structure bring nothing new from the point of view
of integrability and on the other side, for each value of $\xi$ one obtains
new rates which, when positive, give a spectrum with a non-negative real
part.
\section{The spectrum in the one-hole subset. Closed ring}
\setcounter{equation}{0}
$\quad {}$
Our aim is to find the spectrum of the Hamiltonian $H$ given by Eq. (3.10)
in the one-hole sector. This implies solving the equations (3.11) - (3.13)
in the case of the closed ring. For the special case of the left-right
symmetric coagulation-decoagulation model:
\begin{equation}
a_L = a_R = a ; d_L = d_R = d ; b_{R, L} = c = 0
\end{equation}
or
\begin{equation}
\alpha_R = \alpha_L = a + d \; ; \; \gamma_R = \gamma_L = 2 a + d \; , \;
\delta = 0
\end{equation}
the one-hole probability equations were solved for both the ring and the open
chain case \cite{I}. The spectrum is that of free fermions. One can easily show
that in this particular case one can use a similarity transformation in
order to bring the Hamiltonian given by (3.10) into the form (3.4) with
$\eta = (1 + d)^{1/2} $ (we choose $a = 1$). The Glauber model \cite{C} which
again is described by free fermions \cite{SI} corresponds to the choice:
\begin{equation}
a_L = a_R = a \; , \; b_L = b_R = - d \; ; \; d_R = d_L = d \; , \; c = 2d
\end{equation}
or
\begin{equation}
\alpha_R = \alpha_L = a \; , \; \gamma_R = \gamma_L = d \; , \; \delta = 0
\end{equation}

As mentioned already, the Glauber model was solved in the framework of the
one-spin-string problem. One other special case was solved in Refs. \cite{J},
\cite{LU}, which again is a free fermionic case:
\begin{equation}
a_L \neq a_R \; , \; b_{R , L} = d_{R , L} = c = 0
\end{equation}
or
\begin{equation}
a_L = \alpha_L \; , \; a_R = \alpha_R \; , \; \gamma_R = \gamma_L = a_L +
a_R \; , \; \delta = 0
\end{equation}

We now consider the general case. It is convenient to make a change of
variables:
\begin{equation}
K (j , j + n - 1) = R ( n ; j + \frac{n - 1}{2}) = R (n ; p)
\end{equation}
which implies to consider the length of the hole and its center as
variables ($p$ is integer or half-integer). We next take the Fourier
transform of $ R (n ; p )$ :
\begin{equation}
 R ( n , p ) = \sum^{L - 1}_{q = 0 } \;
e^{- \frac{2 \pi i q p}{L}} \; Q ( n ; q )
\end{equation}
and get
\begin{eqnarray}
\frac{1}{D ( q)}
\left\{
\frac{d Q ( n ; q)}{dt} \; + [ (\gamma - \delta) + n \delta ] Q ( n ; q )
\right\}  =  u ( q ) Q ( n - 1 ; q )
\nonumber \\
 + u^{-1} ( q) Q ( n + 1 ; q ) \qquad (q = 0 , 1 , \cdots L - 1)
\end{eqnarray}
with the boundary conditions
\begin{eqnarray}
Q ( 0 ; q ) = Q ( L ; q ) = 0 \quad (q \neq 0)
\nonumber \\
Q ( 0 ; 0 ) = 1 \; ; \; Q ( L ; 0 ) = A e^{- L \delta t}
\end{eqnarray}
where
\begin{equation}
u ( q) = \left[ \right.
\frac{a_R + a_L + i ( a_R - a_L ) tg \frac{\pi q}{L}}{\alpha_R + \alpha_L +
i (\alpha_R - \alpha_L) tg \frac{\pi q}{L}}
\left. \right]^{1/2}
\end{equation}
\begin{eqnarray}
D(q) = \left[ (a_R + a_L ) (\alpha_R + \alpha_L) \right]^{1/2} \; \left[
\cos^2 \frac{\pi q}{L} \qquad \qquad \qquad \qquad
\right.
\nonumber \\
- \left( \frac{a_R - a_L}{a_R + a_L} \right)
\left( \frac{\alpha_R - \alpha_L}{\alpha_R + \alpha_L} \right)
\sin^2 \frac{\pi q}{L}
\left.
+ \frac{i}{2} \sin \frac{2 \pi q}{L}
\left( \frac{\alpha_R - \alpha_L}{\alpha_R + \alpha_L} +
\frac{a_R - a_L}{a_R + a_L} \right)
\right]^{1/2}
\end{eqnarray}
and $A$ is an arbitrary constant. The boundary conditions (4.10) for the
$q = 0 $ case give an inhomogenous system of differential equations. Since
in the present paper we are interested in the spectrum only, we
consider only the homogenous equations and take
\begin{equation}
Q^{\hom} ( n ; q ) = u^n ( q) F ( n , q ) e^{- \lambda t}
\end{equation}
where $\lambda$ are the eigenvalues. The $F ( n ; q )$ satisfy the equation:
\begin{equation}
(E - n) F (n , q) = V ( F ( n + 1 ; q ) + F ( n - 1 ; q ))
\end{equation}
where
\begin{equation}
F (0 ; q) = F (L ; q) = 0
\end{equation}
and
\begin{equation}
E = \frac{\lambda - \gamma}{\delta} + 1 \; ; \; \; V = - \frac{D (q)}{\delta}
\end{equation}
The Equation (4.14) is already known. It appears in the energy spectrum
problem of an electron in a finite one-dimensional crystal in an uniform
electric field \cite{SG}, \cite{MS}. The electric potential ${\cal E} x$ on the
lattice gives
a contribution ${\cal E} n$ on the lattice which appears in the left-hand side
of
Eq. (4.14). (We have divided by ${\cal E}$ the expression of the corresponding
Hamiltonian). For real $D (q)$ the energy levels form the so-called
Wannier-Stark ladder which has been investigated in great detail. Eq. (4.14)
also appears in the problem of the diagonalization of the
$XX$ Hamiltonian in a $Z$ field whose strength is position dependent \cite{RS}:
\begin{equation}
H = - \sum^{N}_{k = - N}
\left[ \sigma^x_k \sigma^x_{k + 1} + \sigma^y_k \sigma^y_{k + 1} + \right.
\left. ( h + R k ) \sigma^z_k \right]
\end{equation}
where $\sigma^x , \sigma^y$ and $\sigma^z$ are Pauli matrices and $h$ and
$R$ are constants. We will shortly present the solution of Eq. (4.14)
as presented
in Ref. \cite{SG}, it is given in terms of Lommel functions \cite{WAT}:
\begin{equation}
R_{m , \mu} (z) = \sum^{\leq \frac{m}{2}}_{n = 0}
\frac{(- 1)^n (m - n)!}{n! (m - 2n)!} \; \;
\frac{\Gamma (\mu + m - n)}{\Gamma (\mu + n )}
\left( \frac{z}{2} \right)^{-m + 2n}
\end{equation}
which satisfy the identity
\begin{equation}
R_{m , \mu} (z) = (-1)^m  R_{m , - \mu - m + 1} (z)
\end{equation}
The eigenvalues $E$ of the equation (4.14) are given by the equation:
\begin{equation}
R_{L - 1, 1 - E} \; \; (2 V (q)) = 0
\end{equation}
There are $(L - 1)$ solutions to the equation (4.20). To each of these
solution corresponds an eigenfunction:
\begin{equation}
F (n, q ) = R_{n - 1 , 1 - E} \; (2V (q)) \; F (1 , q) \; ; \;
( n = 2, \ldots , L - 1 )
\end{equation}
Because of the identity (4.19) if $E$ is an eigenvalue so is $\tilde{E}$ where
\begin{equation}
\tilde{E} = - E + L
\end{equation}
This implies that the eigenvalues $E$ are distributed symmetric around
$\frac{L}{2}$. For $L$ even, $\frac{L}{2}$ is an eigenvalue. For various
properties of the zeroes of the Lommel polynomials see Ref. \cite{WAT}. To the
eigenvalues obtained from Eqs. (4.16) and (4.20), one has to add the
eigenvalue $\lambda = L  \delta$ obtained directly from Eq. (4.10) for
$q = 0$.
The general character of the spectrum for real $V$ is the following. For not
too large $V$, the eigenvalues in the center are equidistant, forming a
simple ladder with unit spacing which results from the l. h. s. of Eq.
(4.14). At the upper and lower end of the spectrum the spacing increases. In
this region the $V$-term in the equation becomes important.

In the limit of large $V$ or small $\delta$ (i. e. for vanishing field in
the electronic problem) one comes back to a simple tight-binding model with
eigenvalues
\begin{equation}
\frac{E}{V} =  2 \cos \frac{\pi k}{L} \; \; ( k = 1 , \ldots L - 1)
\end{equation}
and the corresponding eigenfunction is
\begin{equation}
F ( n , q ) = \frac{\sin ( \frac{n \pi k}{L})}{\sin \frac{\pi k}{L}} \;
F ( 1, q )
\end{equation}
This implies that for $\delta = 0$, the eigenvalues of the Hamiltonian are
\begin{equation}
\lambda ( q ) = \gamma - 2 \; D ( q ) \cos \frac{\pi k}{L} \quad
( k = 1 , 2 , \ldots L - 1 ; q = 0 , 1, .  L - 1 )
\end{equation}
to which we have to add $\lambda = 0$, obtained directly from Eq. (4.10) in
the case $q = 0$. From the expression (4.12) of $D (q)$ we can obtain the
conditions under which the spectrum is that of free fermions. The energy
levels should correspond to two-fermionic excitations (see Sec. 3). If
\begin{equation}
\frac{a_R}{a_L} = \frac{\alpha_R}{\alpha_L}
\end{equation}
one obtains
$$
\lambda (q) = \gamma + G (q + k) + G(q - k) \hspace{7cm} ({\rm 4.27.a})
$$
where
$$
G (k) = -
\left[ (a_R + a_L) (\alpha_R + \alpha_L) \right]^{1/2} \;
\left( \cos \frac{\pi k}{L} + i \frac{a_R -a_L}{a_R + a_L} \sin
\frac{\pi k}{L} \right)
\hspace{1cm} ({\rm 4.27.b})
$$
which indeed looks like a two fermions excitation one of momentum
$(q + k )$, the other of momentum $(q - k)$ with a total momentum $q$. The
known examples (4.2), (4.4) and (4.6) satisfy the condition (4.26). If
$a_R/a_L \neq \alpha_R/\alpha_L$ and if $\delta = 0$ the spectrum is not
given by free fermions, nevertheless the spectrum and wave-functions have a
simple expression.

We have checked on small chains that if the condition (4.26) is fulfilled
the spectrum of
the Hamiltonian (not only the one-hole subset) is indeed given by free
fermions. We didn't look for the similarity transformation which would bring
the Hamiltonian to the corresponding form.

The problem of the open chain, also probably solvable, was not considered yet.
We have only looked at small chains for the simple case where
$b_R = b_L = c = 0$. The levels are simply or doubly degenerate (the $E = 0$
level is always doubly degenerate). This is in contrast to the free
fermionic picture where all levels should be double degenerate (see Sec.
3). This is to be expected as long as the condition (4.26) is not satisfied.
The surprise is that if one takes
\begin{equation}
\setcounter{equation}{28}
\frac{a_R}{a_L} = \frac{\alpha_R}{\alpha_L} \neq 1
\end{equation}
although the periodic chain is "fermionic" this is not the case for the
open chain.

Finally, we didn't touch the other sectors of the Hamiltonian (more than one
hole). We will return to this problem in Sec. 7.

\section{The spectra in the continuum limit (closed ring)}
\setcounter{equation}{0}
$\quad {} $
It is interesting to consider the continuum limit of the spectra. We first
take the case $\delta = 0 $. Then Eq. (4.14) becomes
\begin{eqnarray}
(\frac{E}{V} - 2) F (n, q )
& = & F ( n + 1 ; q ) + F ( n - 1 ; q ) - 2 F ( n ; q ) \nonumber \\
& \approx & \frac{d^2 F ( n ; q )}{d n^2}
\end{eqnarray}
with the boundary condition
\begin{equation}
F ( 0 ; q ) = F ( L ; q ) = 0
\end{equation}
which implies
\begin{equation}
\frac{E}{V} - 2 = - \frac{\pi^2 k^2}{L^2} \quad ( k = 1, \ldots )
\end{equation}
and using Eq. (4.16):
\begin{equation}
\lambda = \gamma - 2 D (q) + D (q) \frac{\pi^2 k^2}{L^2} \qquad
( k = 1, 2, \ldots ; q = 0 , \ldots )
\end{equation}
where (see Eq. (4.12))
\begin{equation}
D ( q ) = \sqrt{(a_R + a_L) (\alpha_R + \alpha_L)}
\left[ 1 + \frac{i \pi q}{2 L} \right.
\left( \frac{\alpha_R - \alpha_L}{\alpha_R + \alpha_L} \right.
\left. + \frac{a_R - a_L}{a_R + a_L} \right)
\left. + 0 \left( \frac{q^2}{L^2} \right)  \right]
\end{equation}
To Eq. (5.4) we have to add the value $\lambda = 0 $ for $q = 0$ (keep in
mind Eq. (4.10)). The spectrum given by Eq. (5.4) is massive if
$\gamma - 2 D (0) > 0$, massless if $\gamma = 2 D (0)$ with a quadratic
dispersion relation in the real part and a linear one in the imaginary part.

We now consider the case $\delta \neq 0$. Using Eqs. (4.14) and (4.16), we
get in the continuum:
\begin{equation}
\frac{d^2 F (n, q)}{d n^2} = (a + b n ) F (n, q )
\end{equation}
where
\begin{equation}
a = \frac{\gamma - \delta - \lambda}{D (q)} - 2 \; \; ; \;
b = \frac{\delta}{D (q)}
\end{equation}
with
\begin{equation}
F ( 0 ; q ) = F ( L ; q ) = 0
\end{equation}
If $Ai (z)$ and $Bi (z)$ are two independent solutions of the Airy equation
\cite{AS}:
\begin{equation}
\frac{d^2 w}{d z^2} = z w
\end{equation}
the general solution of Eq. (5.6) reads
\begin{equation}
F ( n, q) = C_A Ai ( a \; b^{-2/3} + b^{1/3} n) + C_B Bi ( a \; b^{- 2/3} +
b^{1/3} n )
\end{equation}
where $C_A$ and $C_B$ are arbitrary constants.

The eigenvalues $\lambda$ (contained in the constant $a$) are determined
from the boundary conditions (5.8). These equations have to be solved
numerically. In the infinite volume limit ($L \to \infty$) the whole picture
simplifies. Since the Airy function $Bi (z)$ diverges for large z, we are left
with the function $Ai$ only. In order to satisfy the condition
\begin{equation}
F ( 0, q) = 0
\end{equation}
$a \; b^{-2/3}$ has to be a zero of $Ai (z)$. It is known \cite{AS} that the
zeroes
of $Ai (z)$ are all on the negative real axis in the complex $z$ phase
(they are tabulated in \cite{AS}), we give the first two values:
$c_1 = - 2. 338 \; , \; c_2 = - 4.087$. This implies that
\begin{equation}
\lambda_n = \gamma - \delta - 2 D (q) + \mid c_n \mid D^{1/3}_{(q)} \; \;
\delta^{2/3}
\end{equation}
and thus the spectrum in the continuum is known. The eigenfunctions are
given by Eq. (5.10) with $C_B = 0$ and $a$ fixed by Eq. (5.12). To the
eigenvalues
(5.12) one has to add for $q = 0$ the eigenvalue $\lambda = 0$ that one gets
from (4.10) for $L$ infinity. At least in the domain where
$a_{R,L} , b_{R , L} , d_{R , L}$ and $c$ are positive the spectrum has to be
massive. In this case one can use the stochastic interpretation of the
Hamiltonian (3.10). The birth and pair-creation processes (see Eq. (2.4)) come
with independent scales in the evolution operator determining a massive
phase so that as long as $\delta \neq 0$ one is in this phase. This
statement is valid not only for the one-hole subset of the Hamiltonian but
for all the subsets.
\section{The critical dynamics of the one-dimensional Ising model}
\setcounter{equation}{0}
$\quad {}$
As an application of the calculations of spectra presented in the last
section
we consider the dynamics of the one-dimensional Ising model defined by the
equilibrium probability distribution
\begin{equation}
P_{eq} = Z^{-1} \prod^L_{i = 1} \; e ^{\frac{1}{T} \; S_i S_{i + 1}}
\end{equation}
where $Z$ is the partition function:
\begin{equation}
Z = \sum_{S_i = \pm 1} \; \prod^L_{i = 1} e ^{\frac{1}{T} S_i S_{i + 1}}
\end{equation}
The two-point function has the following large distance behaviour
\begin{equation}
< S_i \; S_{i + R} > \sim \; e^{- R/\zeta}
\end{equation}
where the correlation lenght $\zeta$ and the mass $\mu$ are
\begin{equation}
\zeta^{-1} = 2 \mu \; , \; \mu = e^{- \frac{2}{T}}
\end{equation}
We can rewrite (6.2) in a sligthly different form:
\beqn
P_{eq} = Z^{-1} \prod^L_{i = 1}
e ^{\frac{1}{T} (-1){^{(\alpha_i - \alpha_{i + 1})}}} \; \;
= Z^{-1} \prod^L_{i = 1} e ^{\frac{1}{T} (- 1){^{\beta_i}}}
\eeqn
where $ \alpha_i , \beta_i = 0 , 1 $. In the last equality of Eq. (6.5) we
have performed a duality transformation. It was shown in Ref. \cite{A} that the
following relations have to be satisfied by the rates appearing in the master
equation (2.2) in order to have $P_{eq}$ as a steady state solution
(detailed balance):
\begin{equation}
\mu \Gamma^{1 , 0 ; s}_{0 , 0} \; - \Gamma^{0 , 0 ; s}_{1 , 0} \; = \; \mu^2
\Gamma^{1, 1 ; s}_{0, 1} \; - \; \mu \Gamma^{0 , 1 ; s}_{1 1} \; = \;
\Gamma^{0 , 0}_{1 , 1} - \mu^2 \; \Gamma^{1 , 1}_{0 , 0}
\end{equation}
\begin{equation}
\Gamma^{0 , 0 ; a}_{1 , 0} \; + \; \mu \Gamma^{1 , 0 ; a}_{1 , 1} \; + \; \mu^2
\Gamma^{1, 1 ; a}_{1 , 0} \; + \; \mu \Gamma^{0 , 1 ; a}_{0 , 0} \; = \;
2 \mu \Gamma^{1 , 0 ; a}_{0 , 1}
\end{equation}
where
\begin{eqnarray*}
\Gamma^{\alpha , \beta ; s}_{\gamma , \delta} \; = \;
\Gamma^{\alpha , \beta}_{\gamma , \delta} \; + \;
\Gamma^{\beta , \alpha}_{\delta , \gamma} \\
\Gamma^{\alpha , \beta ; a}_{\gamma , \delta} \; = \;
\Gamma^{\alpha , \beta }_{\gamma , \delta} \; - \;
\Gamma^{\beta , \alpha}_{\delta , \gamma}
\end{eqnarray*}
Let us assume that we give a set of rates compatible with Eqs. (6.6) -
(6.7). This defines a Hamiltonian and the physical question one asks is the
behaviour of the energy gap $E_G$ in the thermodynamical limit as $\mu$ goes
to zero (the Ising model in one dimension is critical at $T = 0$). The
inverse of the energy gap gives the time correlation length. As shown in
Ref. \cite{A} in several examples, two scenarios are possible. In the first, as
$\mu \to 0$, $E_G$ stays finite. In this case we have no critical slowing
down although we take local dynamics. In the second scenario, $E_G$
vanishes like
\begin{equation}
E_G \sim \mu^z
\end{equation}
with $z = 2$ (see also Refs. \cite{SI}, \cite{FOR}, \cite{LAG}). It is our aim
to check if
for rates $\Gamma^{\alpha , \beta}_{\gamma , \delta}$ (3.10) given by the
hole picture or spin-string rates
$\tilde{\Gamma}^{\alpha , \beta}_{\gamma , \delta}$ given by Eq. (3.25) one
can obtain new solutions. Since in some cases (see Eq. (4.20) or (4.27)) the
spectrum is complex it is interesting to see if the equilibrium state can be
reached in an oscillatory mode.

We first consider the holes picture. The Eqs. (6.6), (6.7) together with
positivity give the following solution:
\begin{eqnarray}
b_R = b_L = c = 0 \; ; \; \; a_R = a_L + \eta \mu \nonumber \\
d_R = \mu ( a_L - \eta ) \; ; \; d_L = \mu (a_L + \eta ) + \eta \mu^2
\end{eqnarray}
where $a_L > 0$ and $\eta$ is a parameter satisfying the condition:
\begin{equation}
a_L > \mid \eta \mid
\end{equation}
Notice that $\delta = 0$ (see Eq. (3.17)) and that
\begin{equation}
\frac{a_R}{a_L} \neq \frac{d_R}{d_L}
\end{equation}
which implies that we do not have free fermions (see Eq. (4.26))
thus the model we consider
is new. In order to compute the energy gap, we use Eqs. (5.4) and (5.5) to
get:
\begin{equation}
E_G = \gamma - 2 \sqrt{(a_R + a_L) (\alpha_R + \alpha_L)} \; = \;
\frac{a_L}{2} \mu^2 - \frac{\eta \mu^3}{4}
\end{equation}
where we have used Eqs. (3.17) and (6.9). As we see we get no oscillatory
behaviour, the energy gap vanishes for $\mu \to 0$ again with $z = 2$ (see
Eq. (6.8)) in agreement with the universality hypothesis.

We now consider the rates
$\tilde{\Gamma}^{\alpha , \beta}_{\gamma , \delta}$ given by Eq. (3.25), in
this case we find:
\begin{equation}
c =
\frac{8 \mu^2}{(1 - \mu^2)} a_R \; , \; a_R = a_L , d_R = d_L = - b_R = -
b_L = \frac{c}{2}
\end{equation}
which gives:
\begin{eqnarray}
\tilde{\Gamma}^{00}_{01} = \tilde{\Gamma}^{00}_{10} = \tilde{\Gamma}^{01}_{00}
= \tilde{\Gamma}^{10}_{00} =
\tilde{\Gamma}^{11}_{01} = \tilde{\Gamma}^{11}_{10} = \tilde{\Gamma}^{01}_{11}
= \tilde{\Gamma}^{10}_{00} = 0
\nonumber \\
\tilde{\Gamma}^{10}_{01} = \tilde{\Gamma}^{01}_{10} \; ; \; \;
\tilde{\Gamma}^{10}_{01} +
\tilde{\Gamma}^{01}_{10} = \tilde{\Gamma}^{00}_{11} + \tilde{\Gamma}^{11}_{00}
\end{eqnarray}
Now we have free fermions and get back the Glauber model \cite{C}. Since we
want
$a_R$ to be finite for $\mu \to 0$, we choose:
\begin{equation}
c = \frac{\mu^2 \lambda}{2} \; \quad \; ( \lambda > 0 )
\end{equation}
where $\lambda$ is a constant and we get the known result
\begin{equation}
E_G = \lambda \mu^2
\end{equation}
Our results for $E_G$ were obtained only within a sector of the Hamiltonian
and one can always argue that other subsets of differential equations can
give different results. This argument holds for Eq. (6.12) but not for Eq.
(6.14) since in this case the Hamiltonian can be diagonalized completely
\cite{SI}.
\section{Conclusions}
$\quad {}$
We have shown how to bring a Hamiltonian describing a one-dimensional
quantum chain with two states per site and depending on seven parameters to
a block-diagonal form. The structure of the blocks is reminiscent of a
Hamiltonian described by free fermions. The underlying algebraic structure
of the problem is still a mystery for us. We have diagonalized the first
block in the case of periodic boundary conditions. It turns out that Lommel
polynomials play a fundamental role in the eigenvalues and eigenfunctions
of the Hamiltonian. We have also clarified the continuum limit of the spectra.
The case of the open chain was not considered but probably the method used
in one special case \cite{I} can be extended to the general one. Hopefully
the methods similar to the one used for the Glauber model \cite{T} can be
extended to all the blocks. If this is not possible, we will be left
with an interesting example of partially integrable systems in condensed
matter.

We have paid special attention to various similarity transformations which
connect stochastic observables not only in order to find properties of the
spectra coming from positivity but also in order to understand how the
method used in this paper can be extended to
models with more than two states.

In this paper we didn't do any specific calculations related to
diffusion-reaction processes which would have implied solving also the
nonhomogenous case of the differential equations since our purpose was to find
the eigenvalues and eigenfunctions of the Hamiltonian.

One basic question is if our Hamiltonian cannot be diagonalized using
conventional Bethe-Ansatz methods. The answer is not obvious. We first have
checked if one can't use the Baxterisation programme \cite{JO}. Before starting
it we have checked if the matrices $\check{R}_k$:
\begin{equation}
\check{R}_k = H_k + c \hat{1}
\end{equation}
where $H_k$ is given by Eq. (3.10), $c$ is an arbitrary constant and
$\hat{1}$ the unit matrix, satisfy the braid group relations:
\begin{equation}
\check{R}_k  \check{R}_{k + 1}  \check{R}_k = \check{R}_{k + 1}
\check{R}_k \check{R}_{k + 1}
\end{equation}
The answer is in general negative. If the answer would have been positive we
would have had to look for some conditions to find an associative algebra
for which the baxterization programme can be done.
We have next tried the Reshetikhin
criterium according to which the following relation has to be satisfied
\cite{KU}:
\begin{equation}
\left[ H_k + H_{k + 1} , [ H_k , H_{k + 1}] \right] = X_k - X_{k + 1}
\end{equation}
where $X_k$ is an arbitrary $4 \times 4$ matrix. The answer is again negative
in general. One gets a positive answer only in what we called the
free-fermionic case (the parameters satisfy the condition $\delta = 0$ and
Eq. (4.26)). As it is known the Reshetikin condition is very restrictive
(the Hubbard Hamiltonian does not satisfy it) but the authors' knowledge
about conventional integrability conditions stops here.
\section*{Acknowledgements}
$\quad {}$
U. S. would like to thank K. Penson for continuous guidance and the Procope
Programme for financial support. I. P. and V. R. would
like to thank the Deutsche Forschungsgemeinschaft for financial support. V.
R. would like to thank T. Truong and the Laboratoire de Mod\'eles de
Physique Math\'ematiques of the Tours University for hospitality.
I. P. thanks similarly P. Fulde and the MPI Dresden for hospitality.
We would
like to thank M. Droz for asking a question which lead to the Sec. 6 of this
paper, to H. de Vega for pointing out Ref. \cite{KU} and to V. Privman for
drawing our attention to Ref. \cite{KFS}. We would also like to thank for
discussions F. Essler, V. Fateev, G. v. Gehlen, H. Hinrichsen, G. Sch\"utz,
A. Talapov and B. Wehefritz.

\end{document}